\documentclass[british,english]{paper}
\usepackage[latin9]{inputenc}
\usepackage{babel}
\usepackage{textcomp}
\usepackage{graphicx}
\usepackage[authoryear]{natbib}
\usepackage[unicode=true]
 {hyperref}

\makeatletter

\providecommand{\tabularnewline}{\\}

\makeatother

\begin{document}

\title{Do Google Trend data contain more predictability than price returns?}

\author{Damien Challet$^{1,2}$ and Ahmed Bel Hadj Ayed$^{1}$}

\institution{$^{1}$ Chaire de finance quantitative\\
Laboratoire de mathématiques appliquées aux systèmes\\
École Centrale Paris\\
Grande Voie des Vignes, 92295 Châtenay-Malabry, France\\
~\\
$^{2}$ Encelade Capital SA\\
EPFL Innovation Park, Building C\\
1015 Lausanne, Switzerland}
\maketitle
\begin{abstract}
Using non-linear machine learning methods and a proper backtest procedure,
we critically examine the claim that Google Trends can predict future
price returns. We first review the many potential biases that may
influence backtests with this kind of data positively, the choice
of keywords being by far the greatest culprit. We then argue that
the real question is whether such data contain more predictability
than price returns themselves: our backtest yields a performance of
about 17bps per week which only weakly depends on the kind of data
on which predictors are based, i.e. either past price returns or Google
Trends data, or both.
\end{abstract}

\section{Introduction}

Taking the pulse of society with unprecedented frequency and focus
has become possible thanks to the massive flux of data from on-line
services. As a consequence, such data have been used to predict the
present \citep{choi2012predicting} (called \emph{nowcasting }by \citet{castle2009nowcasting}),
that is, to improve estimates of quantities that are being created
but whose figures are to be revealed at the end of a given period.
The latter include unemployment, travel and consumer confidence figures
\citep{choi2012predicting}, quarterly company earnings (from searches
about their salient products) \citep{da2011search}, GDP estimates
\citep{castle2009nowcasting} and influenza epidemics \citep{ginsberg2008detecting}.

The case of asset prices is of particular interest, for obvious reasons.
It seems natural that the on-line activity of people who have actually
traded is related in some way to contemporaneous price changes. However,
forecasting asset price changes with such data is a much harder task.
The idea is by no means recent (see e.g. \citet{antweiler2004all}).
The literature investigates the mood of traders in forums devoted
to finance \citep{antweiler2004all,rechenthin2013stock}, newspapers
\citep{gerow2011mining}, tweets \citep{bollen2011twitter}, blogs
\citep{gilbert2010widespread}, or a selection of them \citep{mao2011predicting}.
Determining the mood of traders requires however to parse the content
of the posts and to classify them as positive or negative. 

A simpler approach consists in using Google Trends (GT thereafter)
which reports historical search volume interest (SVI) of chosen keywords
and to relate SVIs to financial quantities of interest trading volume,
for instance price volatility or price returns \citep{da2011search,gerow2011mining,wang2012impact_supply_demand_on_stock,bordino2012web,takeda2013google,preis2013quantifyingGT,kristoufek2013can}.
Findings can be summarized as follows: using this kind of data to
predict volume or volatility is relatively easy, but the correlation
with future price returns is much weaker. Incidentally, this matches
the daily experience of practitioners in finance who use price returns
instead of fancy big data.

Here we discuss what can go wrong in every step required to backtest
a trading strategy based on GT data. We then use an industry-grade
backtest system based on non-linear machine learning methods to show
the near-equivalence of the exploitable information content between
SVI and historical price returns. We therefore conclude that price
returns and GT contain about the same amount of predictive information,
at least with the methods we have used and challenge to community
to do any better.

\section{Backtesting a speculative strategy based on Google Trends data}

Price returns are believed to be unpredictable by a sizable fraction
of academics. Unconditional raw asset prices are certainly well described
by suitable random walks that contain no predictability whatsoever.
Our experience as practitioners suggest that predictability is best
found conditionally and that linear regressions are not the most efficient
tools to uncover non-randomness in this context. There is essentially
no linear price return auto-correlation; however some significant
cross-correlation are found (in sample) between changes of SVI and
future price returns. One would be tempted to conclude that GT data
do contain more exploitable information than price returns. 

In our opinion, using such methods prevents one to ask the right question
and to assess properly the predictability content of either type of
data. We propose that one should first build a non-linear prediction
algorithm and then feed it with either past returns, GT data, or both,
and finally compare the respective performance of each case.

Before reporting such comparisons, we review some dangers associated
with the use of GT data for prediction. As the saying goes, prediction
is hard, especially about the future. But prediction about the future
\emph{in} the past is even harder because it often seem easier than
it should. It is prone to many kinds of biases that may significantly
alter its reliability, often positively \citep{freeman1992behind,leinweber2007stupid}.
Most of them are due to the regrettable and possibly inevitable tendency
of the future to creep into the past. Any small leak from the future
may empower an unbiased random strategy into a promising candidate
for speculative trading. Let us now look closely at how this happens
when trying to find predictability in GT data. The procedure goes
as follows:
\begin{enumerate}
\item Choose a set of trading strategies
\item Choose the period of backtest
\item Choose a set of assets
\item Choose a set of keywords
\item Download GT data
\item Choose the timescale of returns
\item Choose parameters 
\item Compute the performance with predictors consisting of GT data only,
price returns only, and both.
\end{enumerate}
The rest of the paper is devoted to discuss each of the above steps.

\subsection{Trading strategies}

This must be done first, since otherwise one would backtest all kinds
of strategies until one stumbles on good-looking strategy.

Academic papers often test and report fixed relationships between
an increase of SVI and future price returns. For instance \citet{preis2013quantifyingGT}
assume that an increase in SVI with respect to its moving average
should be followed by a negative return. The same kind of strategies
is found in \citet{kristoufek2013can} who proposes to build a portfolio
whose asset weights decrease as a function of their respective SVI.
All this is unsatisfactory. There is no reason indeed why a given
relationship should hold for the whole period (they do not, see below)
and for all stocks. For instance it is easy to find two assets with
consistently opposite reactions to SVI changes.

Linear strategies are out for the reasons exposed above. One is then
faced with the problem of choosing a family of strategies that will
not overfit the input: there may be many keyword SVIs and functions
thereof as inputs. We choose therefore to use ensemble learning as
a tool to relate different kinds of information and to avoid in-sample
overfitting as much possible. Note, however, that this is only one
layer of stock selection and investment decision in the backtest system
that one of us has implemented.

\subsection{Period of backtest}

The propensity of academic papers to either stop or start their investigations
in 2008, even those written in 2011 \citep{gerow2011mining}, is intriguing.
\citet{kristoufek2013can} uses the whole available length and clearly
shows that the relationship between SVI and future returns has dramatically
changed in 2008. What this means is that one must properly backtest
a strategy with sliding in and out of sample windows \citep{leinweber2007stupid}.
Computer power used to be an issue, but the advent of very cheap cloud
computing power has solved it.

\subsection{Choice of assets}

Most papers are interested in predicting future price returns of a
set of assets, for instance the components of some index (e.g. a subset
of Russell 3000 \citep{da2011search}, Dow Jones Industrial average
\citep{kristoufek2013can}), while some focus on predicting the index
itself \citep{preis2013quantifyingGT}. We focus here on the components
of the S\&P 100 index. The reason why one should work with many assets
is to profit from the power of the central limit theorem: assuming
that one has on average a small edge on each asset price, this edge
will become apparent much faster than if one invests in a single asset
(e.g. an index) at equal edge.

\subsection{Choice of keywords}

This is a crucial ingredient and the most likely cause of overfitting
because one may introduce information from the future into the past
without even noticing it. A distressing number of papers use keywords
from the future to backtest strategies, for instance \citet{preis2013quantifyingGT,choi2012predicting,janetzko2014using}.
One gross error is to think of the keywords that could have been relevant
in the recent past, for instance \texttt{debt}, \texttt{AIG}, \texttt{crisis},
etc. instead of trying to think of ones which will be relevant. But
a much more subtle error is common: to take a set of keywords that
is vague enough and eternally related to finance, for instance \texttt{finance,}
and to find related keywords with Google Sets \citep{preis2013quantifyingGT,choi2012predicting}.
This service suggests a collection of keywords related to a given
set of keywords and is accessible in a spreadsheet from docs.google.com.
We entered a single keyword, \texttt{finance}, and asked for related
keywords. We did not obtain any fancy keywords (\texttt{restaurant,
color, cancer}, etc.) as in \citet{preis2013quantifyingGT}, but did
find the celebrated keyword \texttt{debt}, among others. The problem
is that one cannot ask Google Sets in 2014 what was related to \texttt{finance}
in 2004. As a consequence, the output of Google Sets introduces information
from the future into a backtest. Since, as far as we know, Google
Sets does not provide a wayback machine, it must not be used at all
to augment one's set of keywords used to backtest a strategy. This
shows that the choice of keywords is a crucial ingredient.

In addition, the use of Google was not stationary during the whole
period, which may introduce significant biases into the backtest results.
Correcting them needs at least a null hypothesis, i.e. a null set
of keywords known before the start of the backtest period. This is
why we collected GT data for 200 common medical conditions/ailments/illnesses,
100 classic cars and 100 all-time best arcade games that we trust
were known before 2004 (cf. appendix A) and applied the strategy described
in \citet{preis2013quantifyingGT} with $k=10$. Table \ref{tab:tstats}
reports the t-statistics (t-stats henceforth) of the best three positive
and negative performances (the latter can be made positive by inverting
the prescription of the strategy, transaction costs permitting) for
each set of keywords, including the one from \citet{preis2013quantifyingGT}.

\begin{table}
\begin{centering}
\begin{tabular}{|c|c|}
\hline 
Ailments & t-stat\tabularnewline
\hline 
\hline 
\texttt{multiple sclerosis} & -2.1\tabularnewline
\hline 
\texttt{muscle cramps } & -1.9\tabularnewline
\hline 
\texttt{premenstrual syndrome } & -1.8\tabularnewline
\hline 
\texttt{alopecia} & 2.2\tabularnewline
\hline 
\texttt{gout} & 2.2\tabularnewline
\hline 
\texttt{bone cancer} & 2.4\tabularnewline
\hline 
\end{tabular}~~%
\begin{tabular}{|c|c|}
\hline 
Classic cars & t-stat\tabularnewline
\hline 
\hline 
\texttt{Chevrolet Impala} & -1.9\tabularnewline
\hline 
\texttt{Triumph 2000} & -1.9\tabularnewline
\hline 
\texttt{Jaguar E-type} & -1.7\tabularnewline
\hline 
\texttt{Iso Grifo} & 1.7\tabularnewline
\hline 
\texttt{Alfa Romeo Spider} & 1.7\tabularnewline
\hline 
\texttt{Shelby GT 500} & 2.4\tabularnewline
\hline 
\end{tabular}
\par\end{centering}

\begin{centering}
~\\

\par\end{centering}

\begin{centering}
\begin{tabular}{|c|c|}
\hline 
Classic arcade games & t-stat\tabularnewline
\hline 
\hline 
\texttt{Moon Buggy} & -2.1\tabularnewline
\hline 
\texttt{Bubbles} & -2.0\tabularnewline
\hline 
\texttt{Rampage} & -1.7\tabularnewline
\hline 
\texttt{Street Fighter} & 2.3\tabularnewline
\hline 
\texttt{Crystal Castles} & 2.4\tabularnewline
\hline 
\texttt{Moon Patrol} & 2.7\tabularnewline
\hline 
\end{tabular}\foreignlanguage{british}{~~}%
\begin{tabular}{|c|c|}
\hline 
\citet{preis2013quantifyingGT} & t-tstat\tabularnewline
\hline 
\hline 
\texttt{labor} & -1.5\tabularnewline
\hline 
\texttt{housing} & -1.2\tabularnewline
\hline 
\texttt{success} & -1.2\tabularnewline
\hline 
\texttt{bonds} & 1.9\tabularnewline
\hline 
\texttt{Nasdaq} & 2.0\tabularnewline
\hline 
\texttt{investment} & 2.0\tabularnewline
\hline 
\end{tabular}
\par\end{centering}

\caption{Keywords and associated t-stats of the performance of a simple strategy
using Google Trends time series to predict \texttt{SPY} \label{tab:tstats}
from Monday close to Friday close prices. Transaction costs set at
2bps}
\end{table}

Our brain is hard-wired to make sense of noise and is very good at
inferring false causality. We let the reader ponder about what (s)he
would have concluded if \texttt{bone cancer} or \texttt{Moon Patrol}
were more finance-related. This table also illustrates that the best
t-stats associated to the keyword set of \citet{preis2013quantifyingGT}
are not significantly different from what one would obtain by chance:
the t-stats reported here being a mostly equivalent to Gaussian variables
for time series longer than, say, 20, one expects 5\% of their absolute
values to be larger that 1.95. One notes that \texttt{debt} is not
among the three best keywords when applied to SPY from Monday to Friday:
its performance is unremarkable and unstable, as shown in more details
below. This issue is discussed in more details in \citet{challet2013predicting}.

\subsection{Google Trends data}

Google Trends data are biased in two ways. First, GT data were not
reliably available before 6 August 2008, being updated randomly every
few months \citep{wiki:GT}. Backtests at previous dates include an
inevitable part of science fiction, but are still useful to calibrate
strategies. 

The second problem is that these data are constantly being revised,
for several reasons. The type of data that GT returns was tweaked
in 2012. It used to be made of real numbers whose normalization was
not completely transparent; it also gave uncertainties on these numbers.
Quite consistently, the numbers themselves would change within the
given error bars every time one would download data for the same keyword.
Nowadays, GT returns integer numbers between 0 and 100, 100 being
the maximum of the time-series and 0 its minimum; small changes of
GT data are therefore hidden by the rounding process (but precision
is about 5\% anyway) and error bars are no more available. This format
change is very significant: for instance, the process of rounding
final decimals of prices sometimes introduces spurious predictability,
which is well known for FX data \citep{NeilPrivateComm}. In the case
of GT data, any new maximum increases the granularity of the data,
thereby making it even less reliable. It is one of the reasons members
of \texttt{quantopedian.com} could not replicate the results of \citep{preis2013quantifyingGT}
before the GT data set was released by the authors \citep{quantopian:GT}.
This problem can be partly solved by downloading data for smaller
overlapping time periods and joining the resulting time series.

\subsection{Price returns resolution}

GT data have a weekly resolution by default; most academic papers
make do with such coarse resolution. Note that one downloads them
trimester by trimester, GT data have a daily resolution. As a somewhat
logic consequence, they try to predict weekly price returns. In our
experience, this is \emph{very }ambitious and predictability will
emerge more easily if one times one's investment, if only for instance
because of day-of-the-week effect \citep{gibbons1981day}.

\subsection{Parameter tuning}

Most trading strategies have tunable parameters. Each set of parameters,
which include keywords, defines one or more trading strategies. Trying
to optimize parameters or keywords is equivalent to data snooping
and is bound to lead to unsatisfactory out-of-sample performance.
When backtest results are presented, it is often impossible for the
reader to know if the results suffer from data snooping. A simple
remedy is not to touch a fraction of historical data when testing
strategies and then using it to assess the consistence of its performance,
but only once (cross-validation) \citep{freeman1992behind}. More
sophisticated remedies include White's reality check \citep{white2000reality}
(see e.g. \citet{sullivan1999data} for an application of this method).
Data snooping is equivalent to having no out-of-sample, even when
backtests are properly done with sliding in- and out-of-sample periods.

\begin{figure}
\centering{}\includegraphics[width=0.6\textwidth]{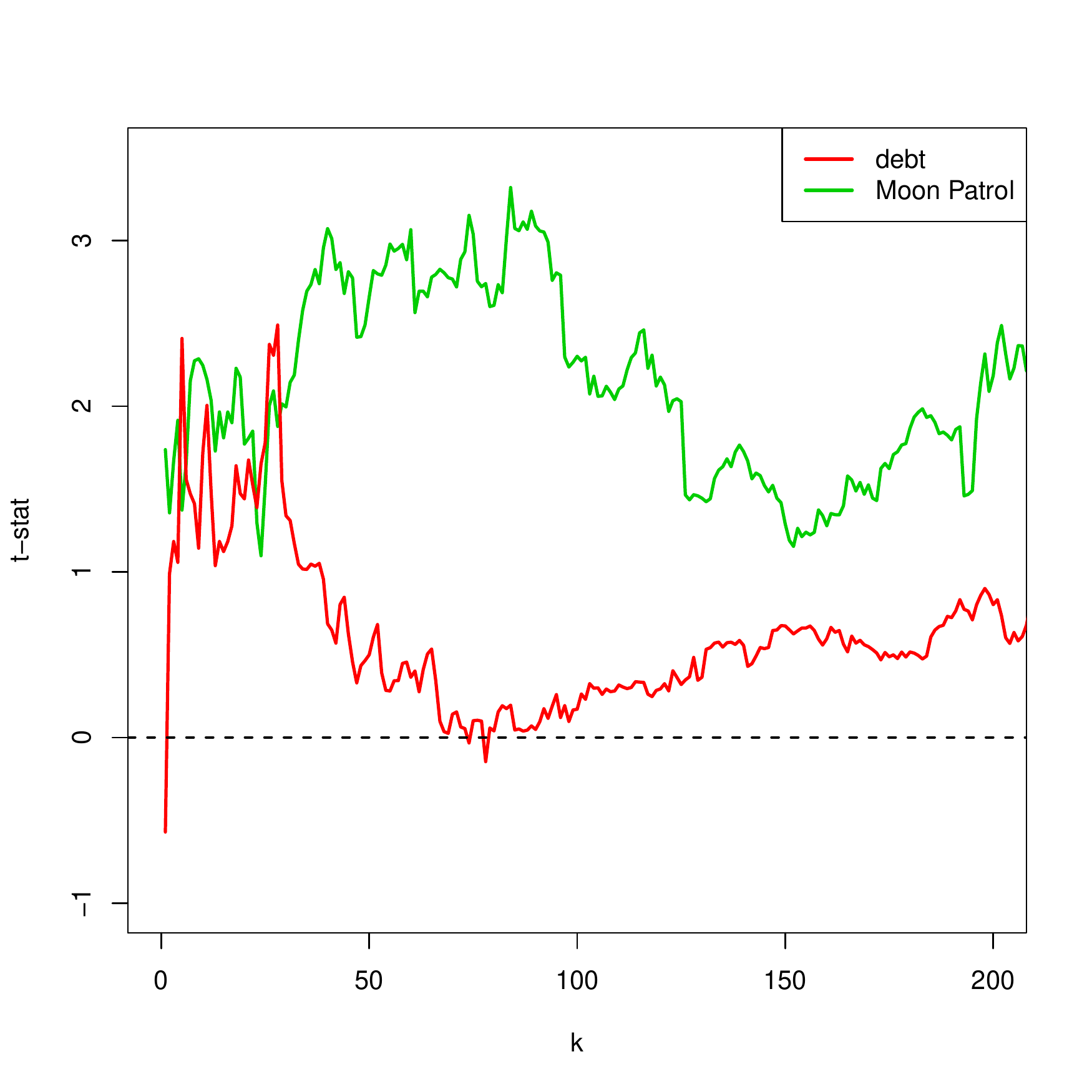}\caption{T-stats of the performance associated with keywords \texttt{debt}
and \texttt{Moon Patrol} versus the length of the moving average $k$.
Transaction costs set to 2bps per transaction.\label{fig:MP-debt-tstat-k}}
\end{figure}

Let us perform some in-sample parameter tuning on the strategy proposed
in \citet{preis2013quantifyingGT}. Figure \ref{fig:MP-debt-tstat-k}
reports the t-stat of the performance associated with the keyword
\texttt{debt} as a function of $k$, the length of the reference simple
moving average. Its sign is relatively robust against changes over
the range of $k\in{2,\cdots,30}$ but its typical value in this interval
is not particularly exceptional. Let us take now the absolute best
keyword from the four sets, \texttt{Moon Patrol}. Both the values
and stability range of its t-stat are much better than those of \texttt{debt}
(see Figure \ref{fig:MP-debt-tstat-k}), but this is entirely due
to pure chance. 

\begin{figure}
\begin{centering}
\includegraphics[width=0.4\textwidth]{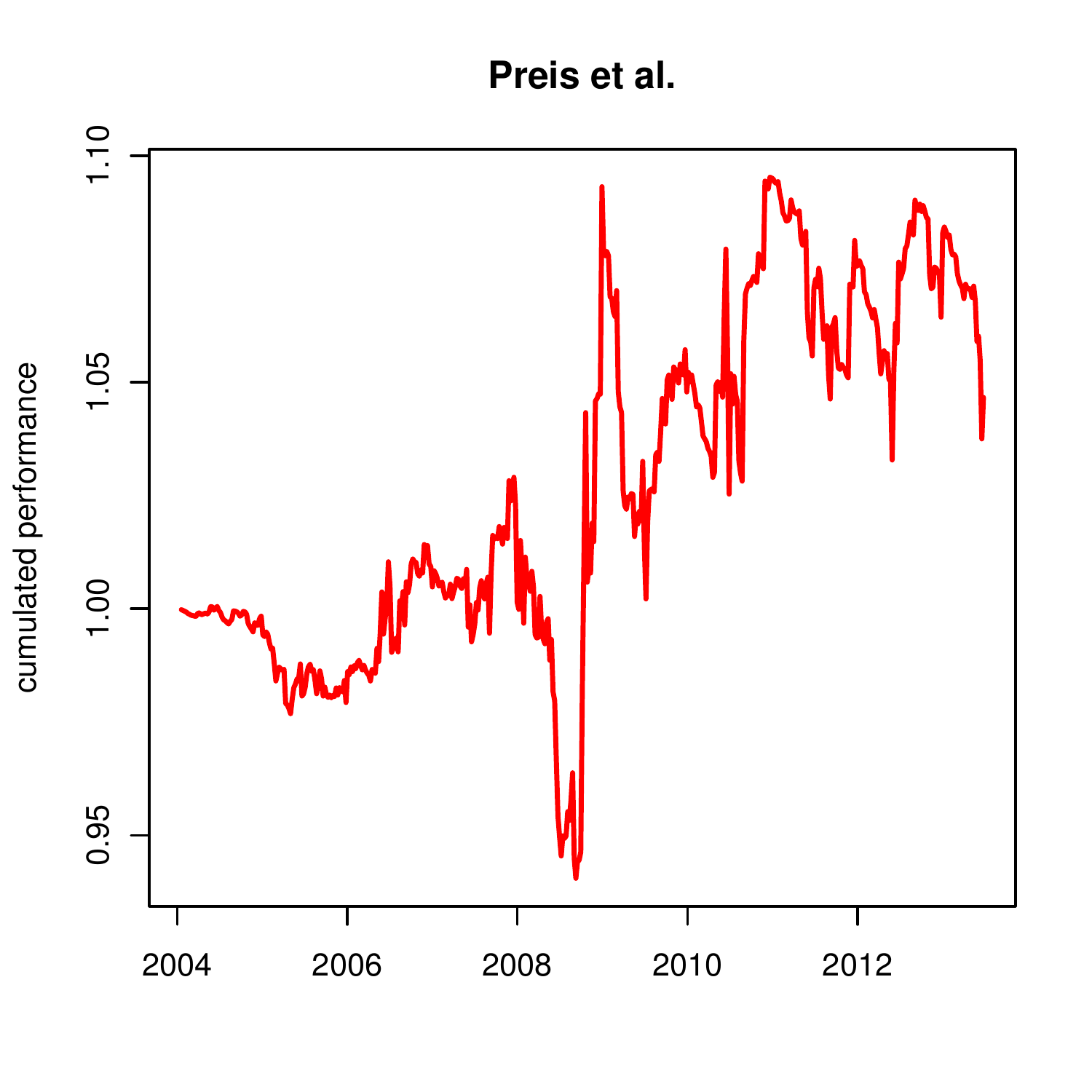}\includegraphics[width=0.4\textwidth]{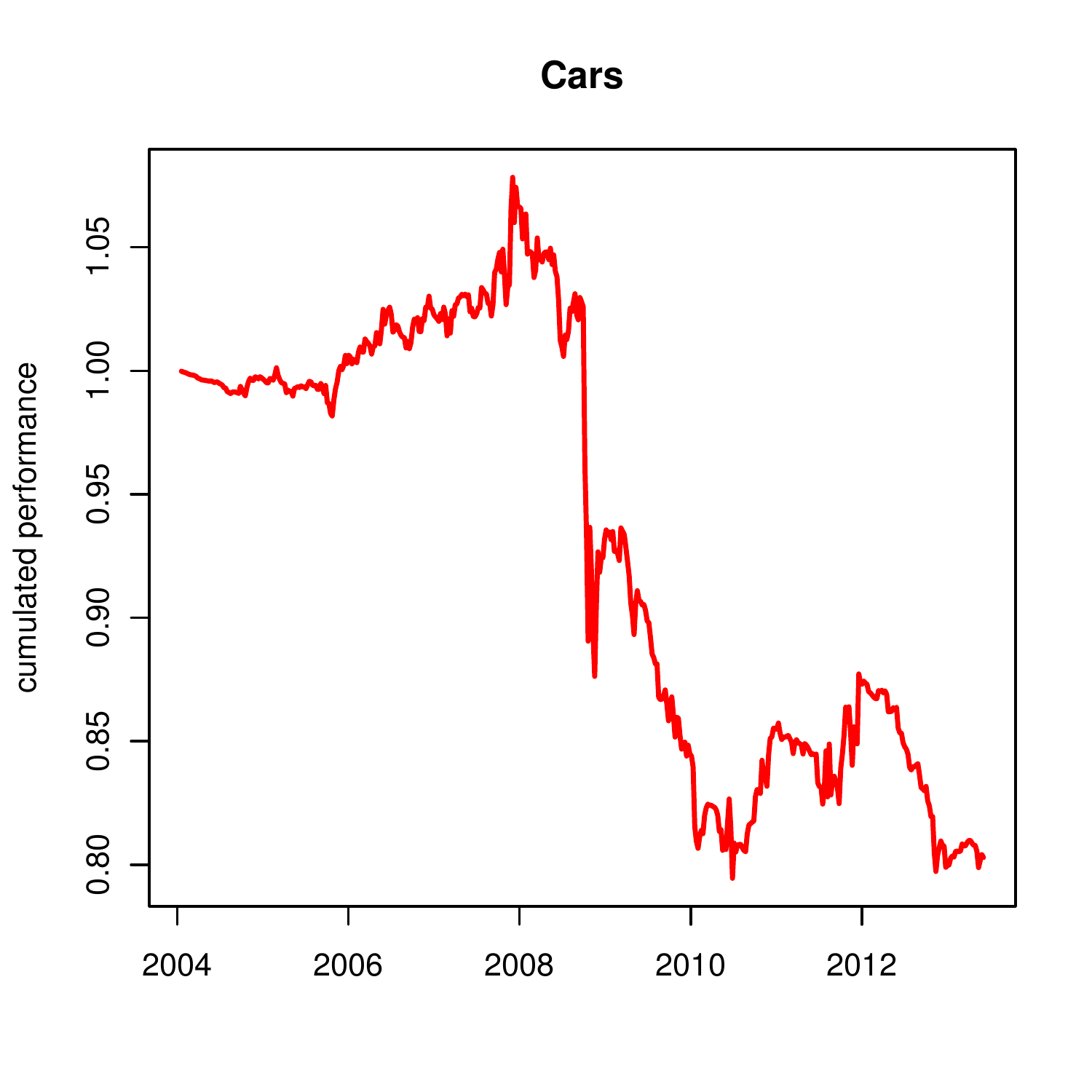}\\

\par\end{centering}

\centering{}\includegraphics[width=0.4\textwidth]{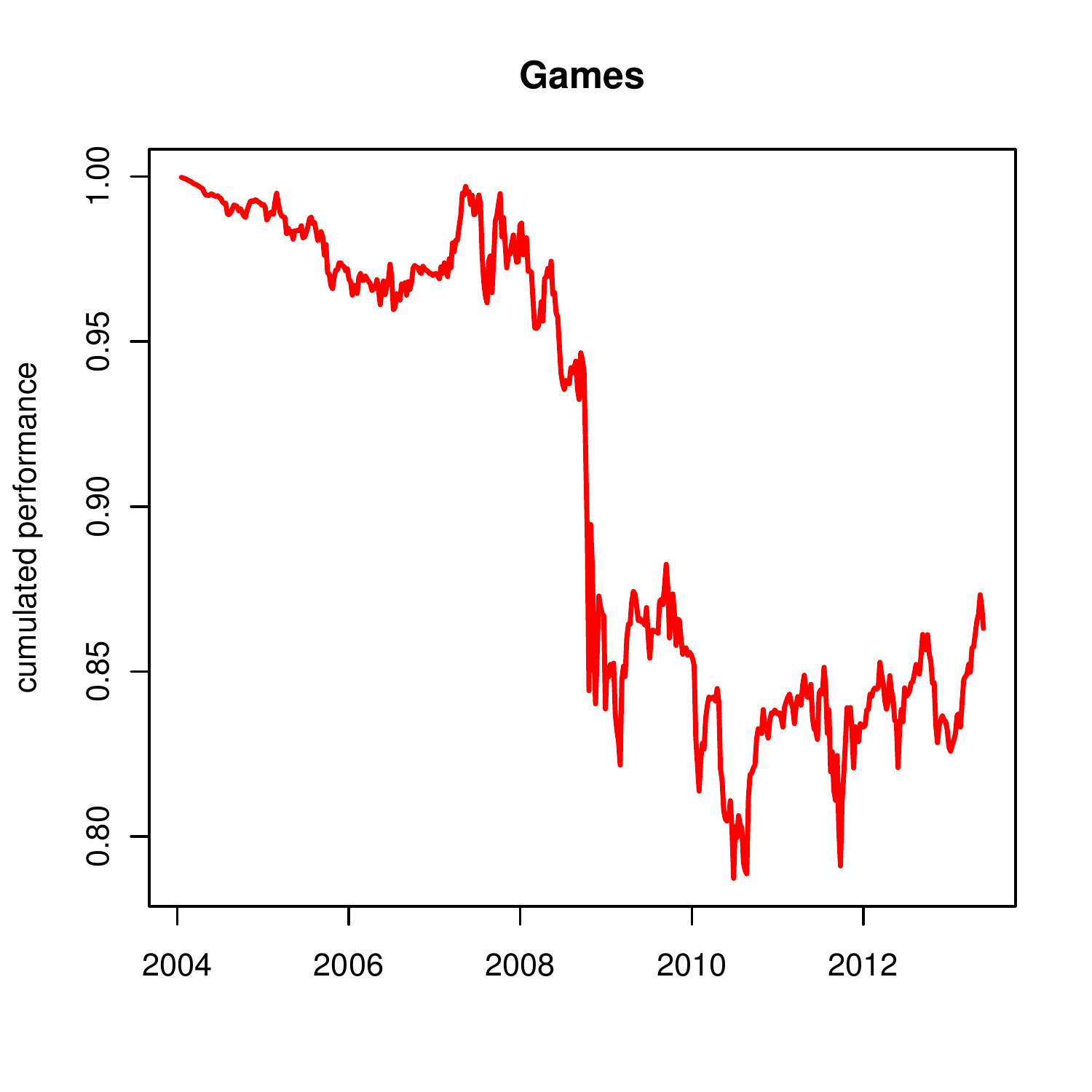}\includegraphics[width=0.4\textwidth]{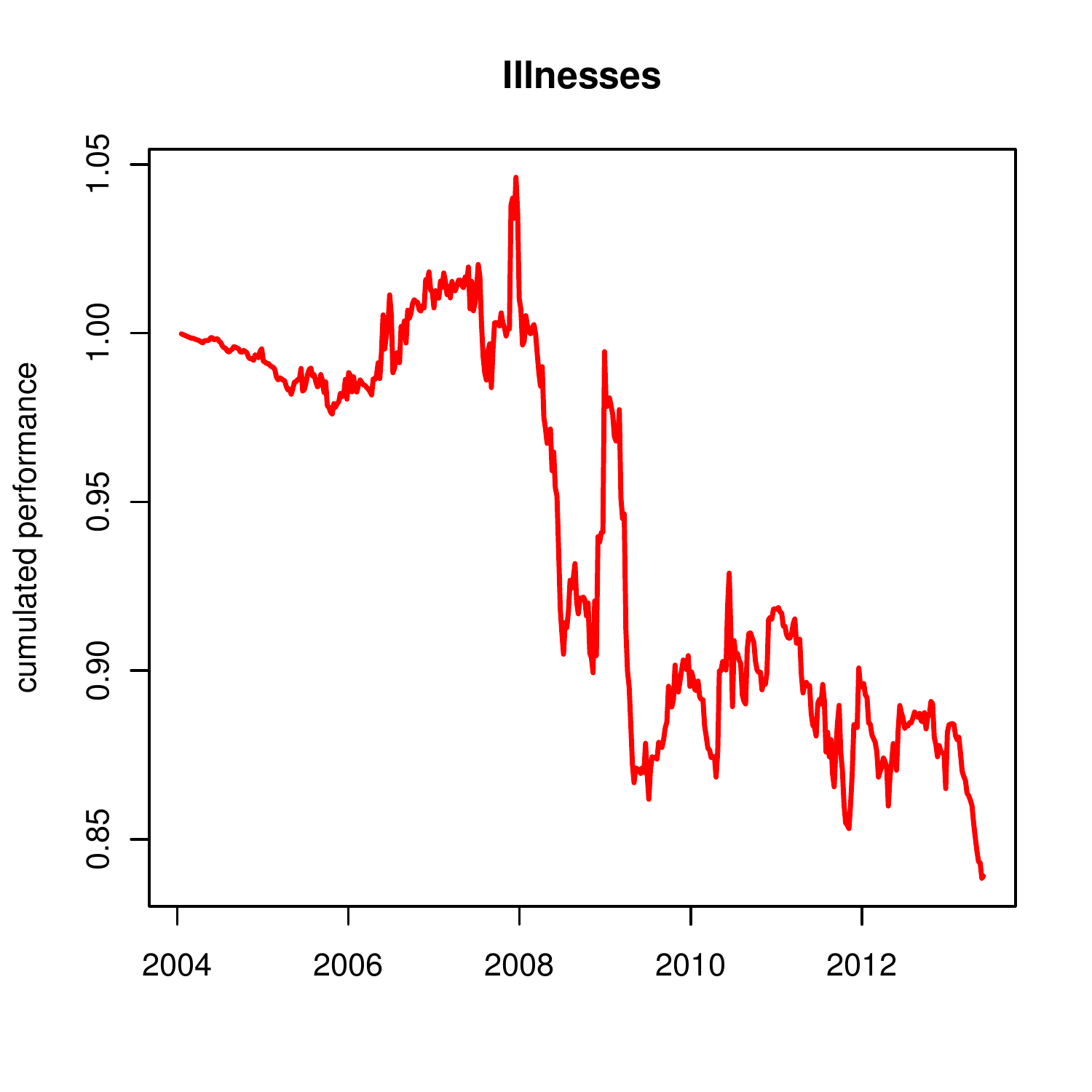}\caption{Cumulated performance associated with the four sets of keywords. Each
transaction costs 2bps.\label{fig:EQW}}
\end{figure}
One solution to avoid parameter overfitting is to average the performance
of a strategy over a reasonable range of parameters. Let us take $k=1,\cdots,100$
for each keyword of each list introduced above. Since all the keywords
act on a single asset, we use for each list an equally weighted scheme
and hence compute the mean position over all keywords and all $k$s.
The resulting cumulated performance net of transaction costs set at
2bps per transaction (which subtracts about 15\% to the performance
computed over the period considered) is reported in Fig.~\ref{fig:EQW}.
It is rather random for random keywords but slightly positive for
the biased keywords of \citet{preis2013quantifyingGT}, which is consistent
with the overall positive bias of t-stats that they report. It is
however not very appealing, with an annualized zero-interest rate
Sharpe ratio of about 0.12 and a t-stat of 0.37, which are far from
being significant. In addition, its performance is flat from 2011
onwards, i.e. out of sample.

\subsection{Compare price returns and GT data as predictors}

\begin{figure}
\includegraphics[width=0.33\textwidth]{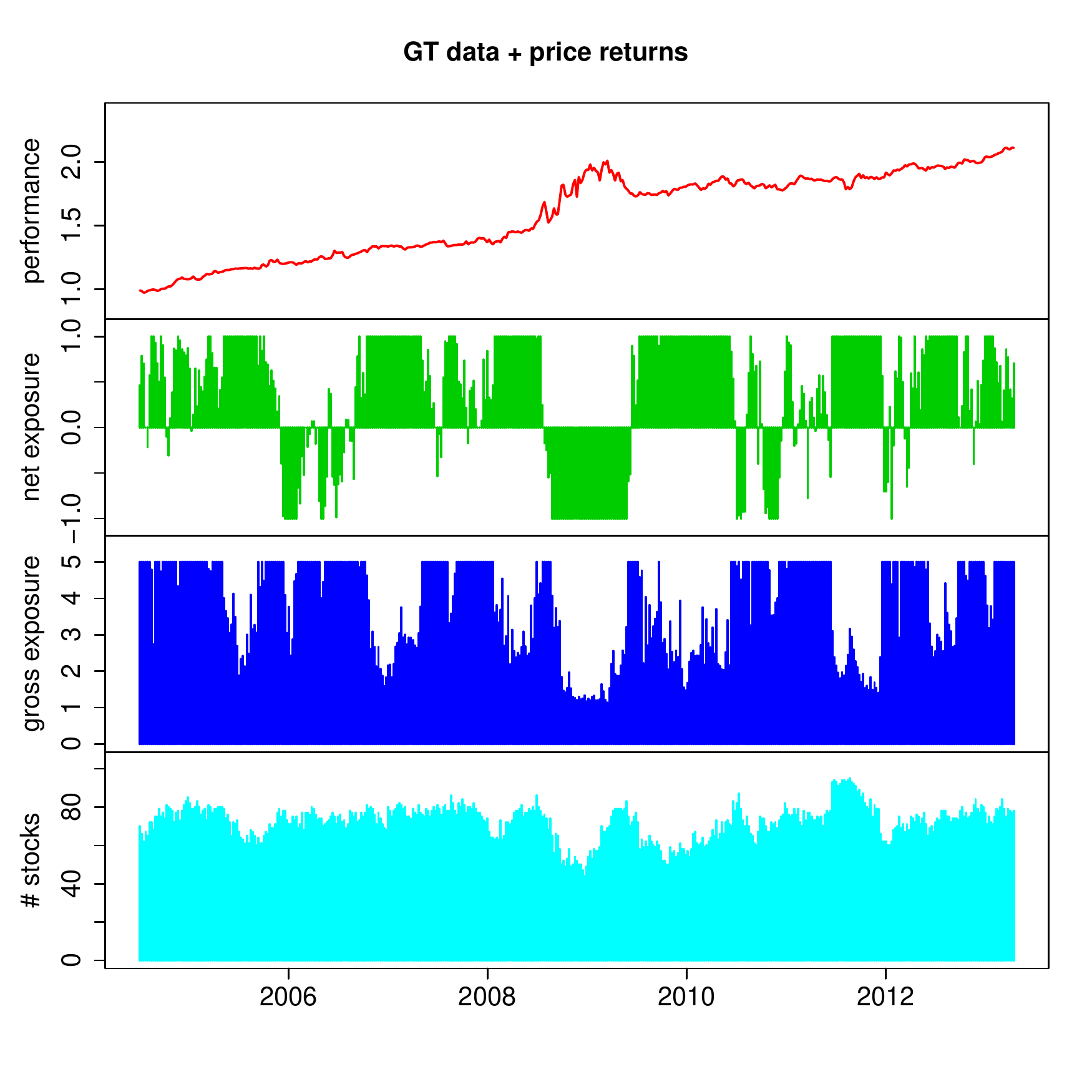}\includegraphics[width=0.33\textwidth]{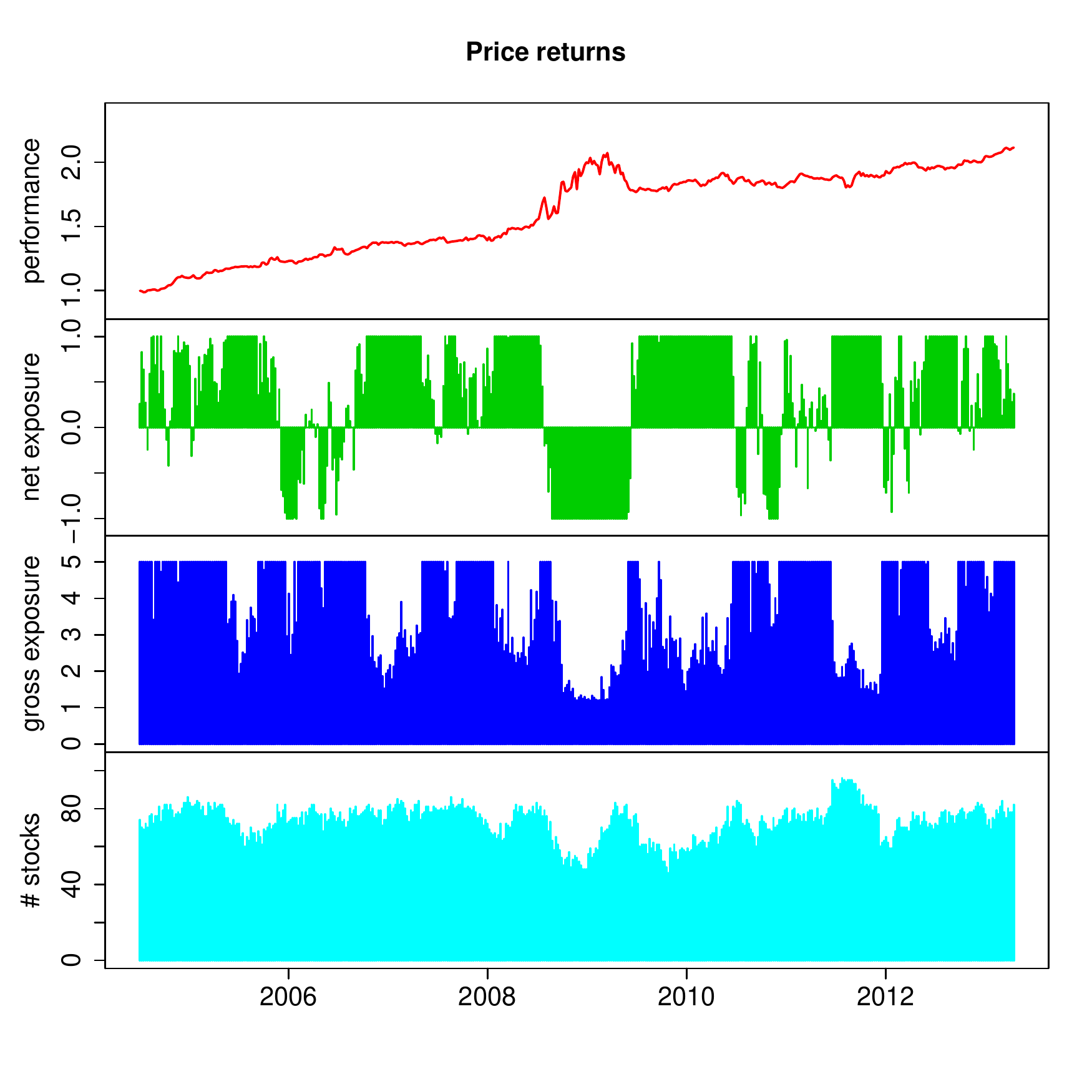}\includegraphics[width=0.33\textwidth]{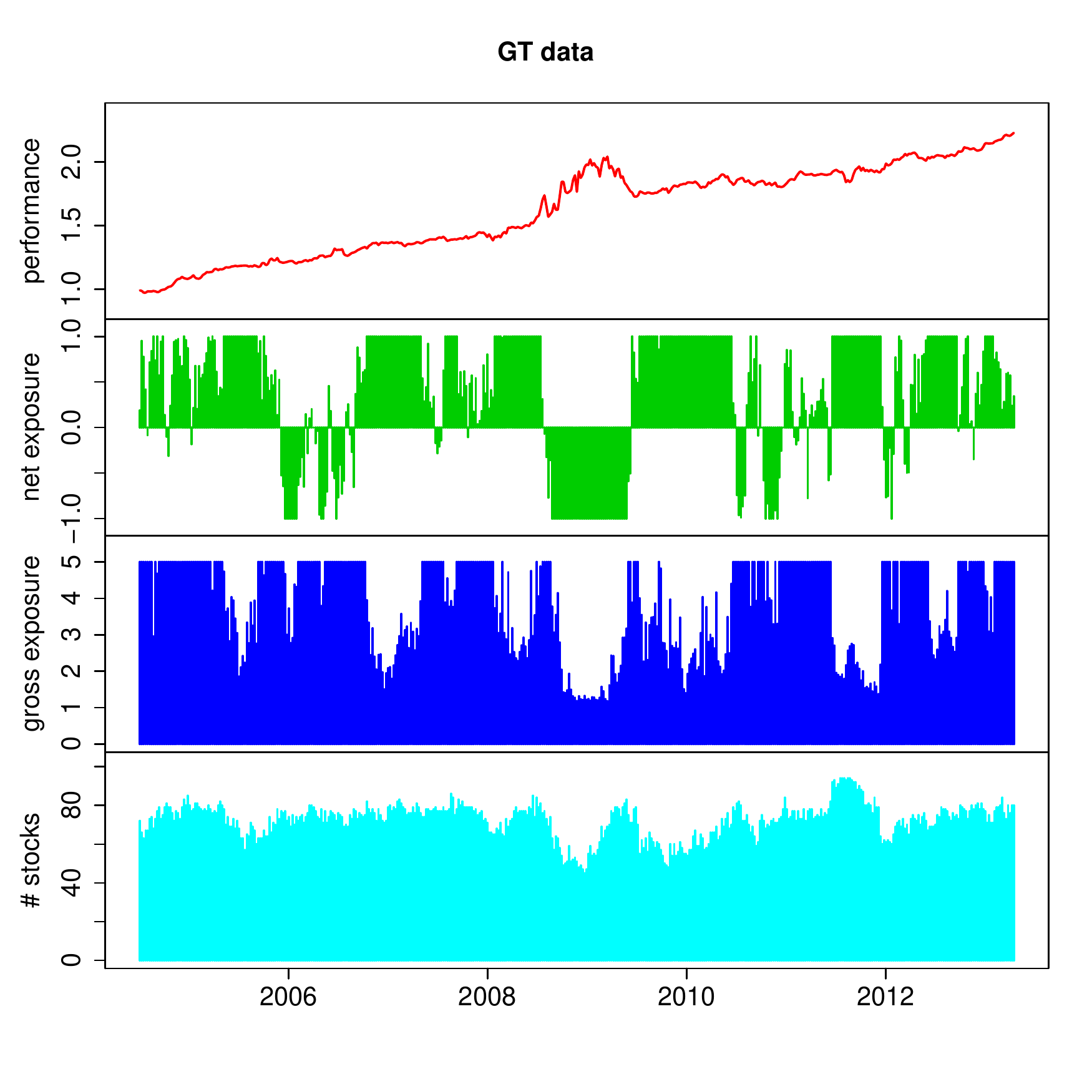}\caption{Cumulated performance, net and gross exposures, and number of stocks
in the portfolio for both GT and returns (left), only GT (middle)
and only price returns (right). Each transaction costs 2bps. \label{fig:backtests_4}}
\end{figure}

\begin{table}
\begin{tabular}{|c|c|c|c|c|}
\hline 
 & weekly return {[}bps{]} & weekly volatility {[}bps{]} & annualized IR & t-stat\tabularnewline
\hline 
\hline 
GT+returns & 17.1 & 134 & 0.92 & 2.73\tabularnewline
\hline 
returns & 16.9 & 133 & 0.92 & 2.72\tabularnewline
\hline 
GT & 18.3 & 134 & 0.99 & 2.93\tabularnewline
\hline 
\end{tabular}\caption{Summary statistics of backtested performance for the three types of
predictors.\label{tab:Summary-statistics}. Each transaction costs
2bps.}
\end{table}
We follow the good idea to choose as keywords company tickers and
names (see e.g. \citet{da2011search,kristoufek2013can}) but add also
other simple, non-overfitting, keywords of our own invention. Weekly
GT data has been downloaded on 2013-04-21. At first, we do not attempt
to predict pure weekly returns but time the investment period. Feeding
our backtest system with GT data and returns yields the leftmost plot
of Figure \ref{fig:backtests_4}: there is some exploitable information
in these data. Calibration window length is about 6 months, which
appears in 2008 and 2009 when the system first learns to take short
positions only and then reverts to long positions. This takes much
time and shows the difficulty one is faced with when calibrating trading
strategies with weekly signals. Summary statistics are reported in
Table~\ref{tab:Summary-statistics}. It is important to be aware
that these backtests are much affected by tool bias, as they use heavy
computational methods and powerful computers that were not available
for most of the backtest period.

\begin{figure}
\centering{}\includegraphics[width=0.6\textwidth]{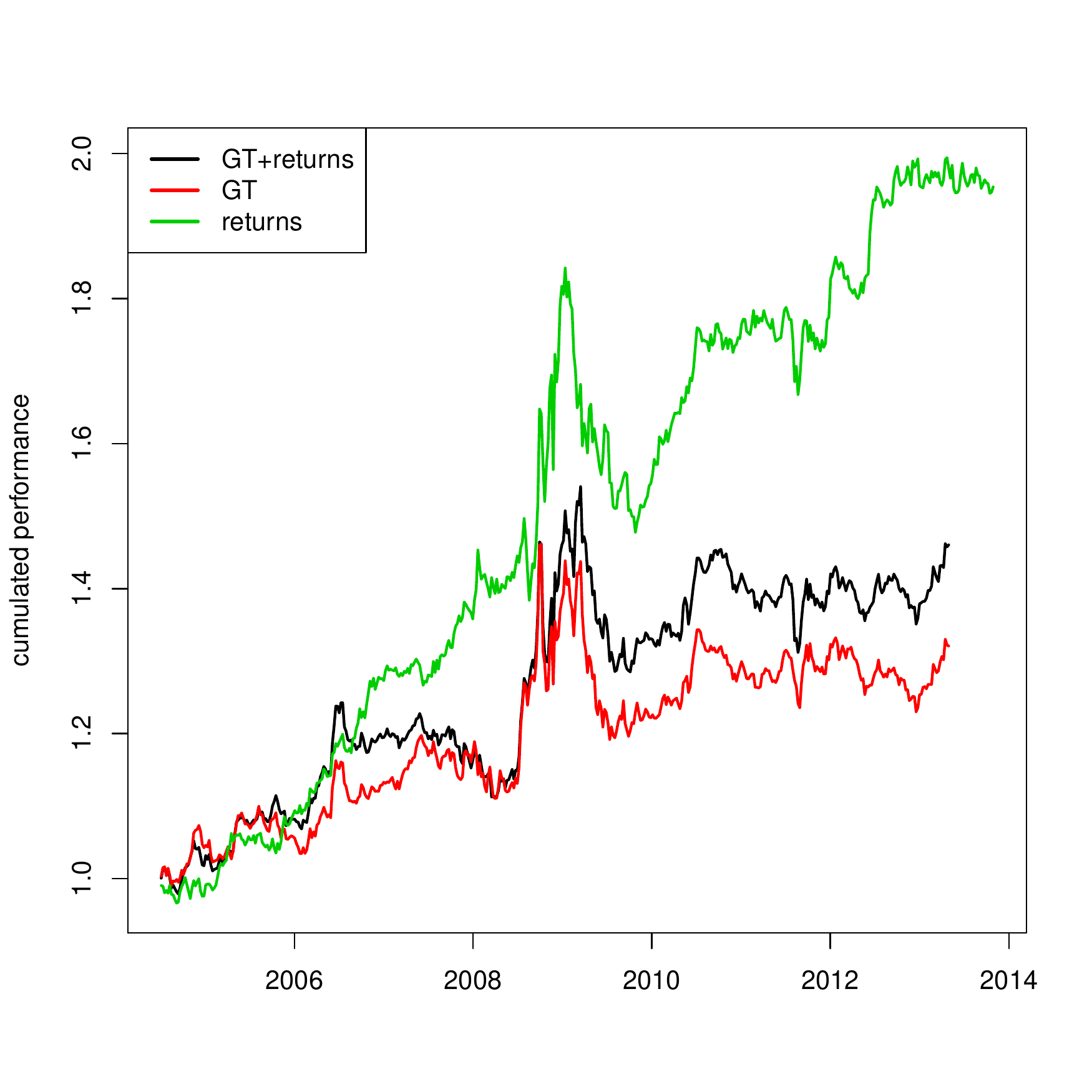}\caption{Cumulated performance both GT and returns, GT only, and returns only
when using binary inputs. Same parameters as Fig.~\ref{fig:backtests_4}.
Transaction costs set at 2bps. \label{fig:binary}}
\end{figure}

Let us now compare the performance of predictors based on GT data
only or past returns only (Fig.~\ref{fig:backtests_4}). We find
essentially the same performance (see Table~\ref{tab:Summary-statistics});
the value of the Wilcox rank sum test p-value is 0.72: they are not
very different. This may be due to the fact that the backtest system
just learns to recognize trends unconditionally, in other words, that
the predictors are simply equally useless. We therefore remove some
information content from the predictors. This is done for example
by computing a rolling median of each predictor; a value of the predictor
is now reduced to a binary number which encodes which side of the
previous median it belongs to. We then use exactly the same backtesting
system as before with the same parameters. The performance associated
to GT data and price returns is now unambiguous (Fig.~\ref{fig:binary}).
The machine learning method used here could exploit less predictability
from GT inputs (at least those we could think of) than from inputs
based on price returns; however, other machine learning methods yield
the opposite result. 

\begin{figure}
\centering{}\includegraphics[width=0.6\textwidth]{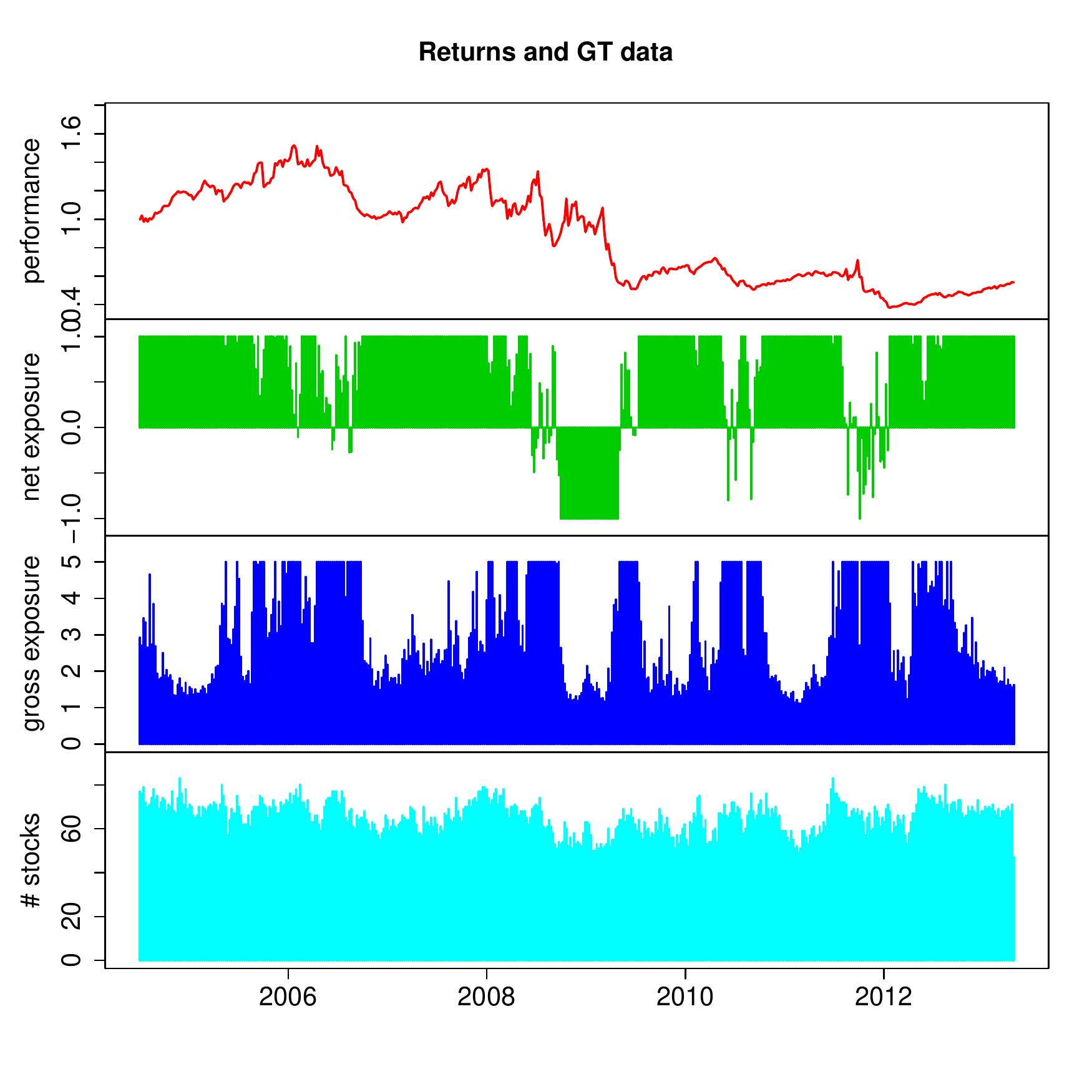}\caption{Cumulated performance, net and gross exposures, and number of stocks
in the portfolio for both GT and returns. Positions may be changed
at the market close of each Monday. No transaction costs.\label{fig:weekly}}
\end{figure}

Finally, Figure \ref{fig:weekly} reports the result of the same method
applied on weekly returns, which shows how hard prediction may be
in this case even without transaction costs.

\section{Discussion}

We have not been able to show that Google Trends data contain more
exploitable information than price returns themselves. Assuming that
this is not due to us not using the right method, our findings suggest
that Google Trends data are equivalent to price return themselves.
They do share indeed many properties with price returns: they are
aggregate signals created by many individuals, they reflect something
related to the underlying assets. In addition both are very noisy:
the uncertainty about GT data, gathered from their previous file format,
is about 5\%. From this point of view, there is nothing miraculous
or ground-breaking about GT data.

We had to use sophisticated non-linear methods coupled with a careful
backtest procedure, which contrasts with the much simpler approaches
usually seen in current academic literature. Why it was needed at
all is probably because it is hard to guess at first what an increase
in SVI means, since it may be related to good news (e.g. higher interest
from potential customers), or bad ones (e.g. worry about the company
itself), or both, or neither. Indeed, such data include too many searches
unrelated to the financial assets for a given keyword, and even many
more unrelated to actual trading. As a consequence, adding another
signal based on the change of the number of news related to a given
asset helps to interpret what a change of SVI means \citep{wang2012impact_supply_demand_on_stock,quant3.0mood}.
Another possibility is to use other sources of data, such as Twitter
or Wikipedia \citep{moat2013quantifying}, which have the invaluable
advantage of being available at a much higher frequency. At any rate,
we challenge the community to show that for a given backtest system,
predictors based on weekly Google Trends data only are able to outperform
predictors based on price that themselves yield about 17bps per week
including 2bps transaction costs.

We acknowledge stimulating discussions with Frédéric Abergel, Marouanne
Anane (Ecole Centrale) and Thierry Bochud (Encelade Capital).

\bibliographystyle{plainnat}
\bibliography{biblio}

\appendix

\section{Keywords}

We have downloaded GT data for the following keywords, without any
manual editing.

\subsection{Illnesses}

Source:\href{http://www.ranker.com/list/list-of-common-diseases-most-common-illnesses/diseases-and-medications-info}{http://www.ranker.com/list/list-of-common-diseases-most-common-illnesses/diseases-and-medications-info},
accessed on 27 May 2013\\
\texttt{\footnotesize AIDS, Acne, Acute bronchitis, Allergy, Alopecia,
Altitude sickness, Alzheimer's disease, Andropause, Anorexia nervosa,
Antisocial personality disorder, Arthritis, Asperger syndrome, Asthma,
Attention deficit hyperactivity disorder, Autism, Avoidant personality
disorder, Back pain, Bad Breath, Bedwetting, Benign prostatic hyperplasia,
Bipolar disorder, Bladder cancer, Bleeding, Body dysmorphic disorder,
Bone cancer, Borderline personality disorder, Bovine spongiform encephalopathy,
Brain Cancer, Brain tumor, Breast cancer, Burns, Bursitis, Cancer,
Canker Sores, Carpal tunnel syndrome, Cervical cancer, Cholesterol,
Chronic Childhood Arthritis, Chronic Obstructive Pulmonary Disease,
Coeliac disease, Colorectal cancer, Conjunctivitis, Cradle cap, Crohn's
disease, Dandruff, Deep vein thrombosis, Dehydration, Dependent personality
disorder, Depression, Diabetes mellitus, Diabetes mellitus type 1,
Diaper rash, Diarrhea, Disabilities, Dissociative identity disorder,
Diverticulitis, Down syndrome, Drug abuse, Dysfunctional uterine bleeding,
Dyslexia, Ear Infections, Ear Problems, Eating Disorders, Eczema,
Edwards syndrome, Endometriosis, Epilepsy, Erectile dysfunction, Eye
Problems, Fibromyalgia, Flu, Fracture, Freckle, Gallbladder Diseases,
Gallstone, Gastroesophageal reflux disease, Generalized Anxiety Disorder,
Genital wart, Glomerulonephritis, Gonorrhoea, Gout, Gum Diseases,
Gynecomastia, HIV, Head Lice, Headache, Hearing impairment, Heart
Disease, Heart failure, Heartburn, Heat Stroke, Heel Pain, Hemorrhoid,
Hepatitis, Herniated Discs, Herpes simplex, Hiatus hernia, Histrionic
personality disorder, Hyperglycemia, Hyperkalemia, Hypertension, Hyperthyroidism,
Hypothyroidism, Infectious Diseases, Infectious mononucleosis, Infertility,
Influenza, Iron deficiency anemia, Irritable Male Syndrome, Irritable
bowel syndrome, Itching, Joint Pain, Juvenile Diabetes, Kidney Disease,
Kidney stone, Leukemia, Liver tumour, Lung cancer, Malaria, Melena,
Memory Loss, Menopause, Mesothelioma, Migraine, Miscarriage, Mucus
In Stool, Multiple sclerosis, Muscle Cramps, Muscle Fatigue, Muscle
Pain, Myocardial infarction, Nail Biting, Narcissistic personality
disorder, Neck Pain, Obesity, Obsessive-compulsive disorder, Osteoarthritis,
Osteomyelitis, Osteoporosis, Ovarian cancer, Pain, Panic attack, Paranoid
personality disorder, Parkinson's disease, Penis Enlargement, Peptic
ulcer, Peripheral artery occlusive disease, Personality disorder,
Pervasive developmental disorder, Peyronie's disease, Phobia, Pneumonia,
Poliomyelitis, Polycystic ovary syndrome, Post-nasal drip, Post-traumatic
stress disorder, Premature birth, Premenstrual syndrome, Propecia,
Prostate cancer, Psoriasis, Reactive attachment disorder, Renal failure,
Restless legs syndrome, Rheumatic fever, Rheumatoid arthritis, Rosacea,
Rotator Cuff, Scabies, Scars, Schizoid personality disorder, Schizophrenia,
Sciatica, Severe acute respiratory syndrome, Sexually transmitted
disease, Sinusitis, Skin Eruptions, Skin cancer, Sleep disorder, Smallpox,
Snoring, Social anxiety disorder, Staph infection, Stomach cancer,
Strep throat, Sudden infant death syndrome, Sunburn, Syphilis, Systemic
lupus erythematosus, Tennis elbow, Termination Of Pregnancy, Testicular
cancer, Tinea, Tooth Decay, Traumatic brain injury, Tuberculosis,
Ulcers, Urinary tract infection, Urticaria, Varicose veins.}{\footnotesize \par}

\subsection{Classic cars}

Source:\href{http://www.ranker.com/crowdranked-list/the-best-1960_s-cars}{http://www.ranker.com/crowdranked-list/the-best-1960\_{}s-cars},
accessed on 27 May 2013\\
\texttt{\footnotesize 1960 Aston Martin DB4 Zagato, 1960 Ford, 1961
Ferrari 250 SWB, 1961 Ferrari 250GT California, 1963 Corvette, 1963
Iso Griffo A3L, 1964 Ferrari 250 GTL (Lusso), 1965 Bizzarrini 5300
Strada, 1965 Ford GT40, 1965 Maserati Mistral, 1965 Shelby Cobra,
1966 Ferrari 365P, 1966 Maserati Ghibli, 1967 Alfa Romeo Stradale,
1967 Ferrari 275 GTB/4, 1967 Shelby Mustang KR500, 1968 Chevrolet
Corvette L88, 1968 DeTomaso Mangusta, 1969 Pontiac Trans Am, 1969
Yenko Chevelle, 57 Chevy, 68 Ferrari 365 GTB/4Daytona Spyder, 69 Yenko
Camaro Z28, AC Cobra, Alfa Romeo Spider, Aston Martin DB5, Austin
Mini Saloon 1959, BMW E9, Buick Riviera, Buick Wildcat, Cane, Chevrolet
Camaro, Chevrolet Chevelle, Chevrolet Impala, Chevy Chevelle, Chrysler
Valiant, Corvette Stingray, Dodge Challenger, Dodge Charger, Dodge
Dart Swinger, Facel Vega Facel II, Ferrari 250, Ferrari 250 GTO, Ferrari
250 GTO, Ferrari 275, Ferrari Daytona, Fiat 500, Ford Corsair, Ford
Cortina, Ford GT40, Ford Mustang, Ford Ranchero, Ford Thunderbird,
Ford Torino, Ford Zephyr MK III, Iso Grifo, Jaguar E-type, Jeep CJ,
Lamborghini Miura, Lamborghini Miura SV, Lincoln Continental, Lotus
Elan, Maserati Ghibli, Mercedes Benz 220SE, Mercedes-Benz 300SL, Mercury
Cougar, Plymouth Barracuda, Pontiac GTO, Porsche 356, Porsche 911,
Porsche 911, Porsche 911 classic, Rambler Classic, Rover 2000, Shelby
Daytona Coupe, Shelby GT350, Shelby GT500, Studebaker Avanti, Sunbeam
Tiger, Toyota 2000GT, Triumph 2000, Vauxhall Velox 1960, Vauxhall
Victor 1963, Wolseley 15/60}{\footnotesize \par}

\subsection{Arcade Games}

Source:\href{"http://www.ranker.com/list/list-of-common-diseases-most-common-illnesses/diseases-and-medications-info}{http://www.ranker.com/list/list-of-common-diseases-most-common-illnesses/diseases-and-medications-info},
accessed on 27 May 2013\\
\texttt{\footnotesize 1942, 1943, 720\textdegree{}, After Burner,
Airwolf, Altered Beast, Arkanoid, Asteroids, Bad Dudes Vs. DragonNinja,
Bagman, Battlezone, Beamrider, Berzerk, Bionic Commando, Bomb Jack,
Breakout, Bubble Bobble, Bubbles, BurgerTime, Centipede, Circus Charlie,
Commando, Crystal Castles, Cyberball, Dangar - Ufo Robo, Defender,
Dig Dug, Donkey Kong, Donkey Kong 3, Donkey Kong Junior, Double Dragon,
Dragon's Lair, E.T. (Atari 2600), Elevator Action, Final Fight, Flashback,
Food Fight, Frogger, Front Line, Galaga, Galaxian, Gauntlet, Geometry
Wars, Gorf, Gorf, Gyruss, Hogan's Alley, Ikari Warriors, Joust, Kangaroo,
Karate Champ, Kid Icarus, Lode Runner, Lunar Lander, Manic Miner,
Mappy, Marble Madness, Mario Bros., Millipede, Miner 2049er, Missile
Command, Moon Buggy, Moon Patrol, Ms. Pac-Man, Naughty Boy, Pac-Man,
Paperboy, Pengo, Pitfall!, Pole Position, Pong, Popeye, Punch-Out!!,
Q{*}bert, Rampage, Red Baron, Robotron: 2084, Rygar: The Legendary
Adventure, Sewer Sam, Snow Bros, Space Invaders, Spy Hunter, Star
Wars, Stargate, Street Fighter, Super Pac-Man, Tempest, Tetris, The
Adventures of Robby Roto!, The Simpsons, Time Pilot, ToeJam \& Earl,
Toki, Track \& Field, Tron, Wizard Of Wor, Xevious }
\end{document}